\documentclass[10pt,letterpaper,english,conference]{IEEEtran}
\usepackage[T1]{fontenc}
\usepackage[latin9]{inputenc}
\usepackage{xcolor}
\usepackage{pdfcolmk}
\usepackage{units}
\usepackage{amsmath}
\usepackage{graphicx}
\usepackage{setspace}
\usepackage[numbers]{natbib}
\PassOptionsToPackage{normalem}{ulem}
\usepackage{ulem}

\makeatletter

\pdfpageheight\paperheight
\pdfpagewidth\paperwidth

\providecolor{lyxadded}{rgb}{0,0,1}
\providecolor{lyxdeleted}{rgb}{1,0,0}

\makeatletter
\IEEEoverridecommandlockouts
\def\ps@headings{%
\def\@oddhead{\mbox{}\scriptsize\rightmark \hfil \thepage}%
\def\@evenhead{\scriptsize\thepage \hfil \leftmark\mbox{}}%
\def\@oddfoot{}%
\def\@evenfoot{}}
\makeatother
\makeatother

\usepackage{babel}
\begin{document}

\title{\textcolor{black}{Effects of MAC Approaches on Non-Monotonic Saturation
with COPE - A Simple Case Study\thanks{This work is sponsored by the Department of Defense under Air Force Contract FA8721-05-C-0002. Opinions, interpretations, recommendations, and conclusions are those of the authors and are not necessarily endorsed by the United States Government. Specifically, this work was supported by Information Systems of ASD(R\&E). Contributions of the Irwin Mark Jacobs and Joan Klein Jacobs Presidential Fellowship have also been critical to the success of this project.}}}

\author{\textcolor{black}{Jason Cloud{*}$^{\dagger}$, Linda Zeger$^{\dagger}$,
Muriel Médard{*}}\\
\textcolor{black}{{*}Research Laboratory of Electronics, Massachusetts
Institute of Technology, Cambridge, MA.}\\
\textcolor{black}{$^{\dagger}$MIT Lincoln Laboratory, Lexington,
MA.}\\
\textcolor{black}{Email: \{jcloud@, zeger@ll., medard@\}mit.edu}}
\maketitle
\begin{abstract}
\begin{onehalfspace}
We construct a simple network model to provide insight into network
design strategies. We show that the model can be used to address various
approaches to network coding, MAC, and multi-packet reception so that
their effects on network throughput can be evaluated. We consider
several topology components which exhibit the same non-monotonic saturation
behavior found within the Katti\textit{ et. al.} COPE experiments.
W\textcolor{black}{e further show that fairness allocation by the
MAC can seriously impact performance and cause this non-monotonic
saturation. Using our model, we develop a MAC that provides }\textit{\textcolor{black}{monotonic}}\textcolor{black}{{}
saturation, higher saturation throughput gains and fairness among
}\textit{\textcolor{black}{flows}}\textcolor{black}{{} rather than }\textit{\textcolor{black}{nodes}}\textcolor{black}{.
The proposed model provides an estimate of the achievable gains for
the cross-layer design of network coding, multi-packet reception,
and MAC showing that super-additive throughput gains on the order
of six times that of routing are possible.}\end{onehalfspace}

\end{abstract}
\IEEEpeerreviewmaketitle

\section{\textcolor{black}{Introduction}}

\textcolor{black}{With the multitude of network technologies available
that increase performance, it is difficult to efficiently integrate
them into a coherent network. We develop a simple model that provides
insight into cross-layer network design, and illustrate its use through
examples involving the combination of the 802.11 medium access control
(MAC), network coding (NC), and multi-packet reception (MPR). Our
model not only provides strategies to integrate various technologies,
but also predicts the achievable gains possible.}
\begin{figure}
\begin{centering}
\textcolor{black}{\includegraphics[width=3.25in]{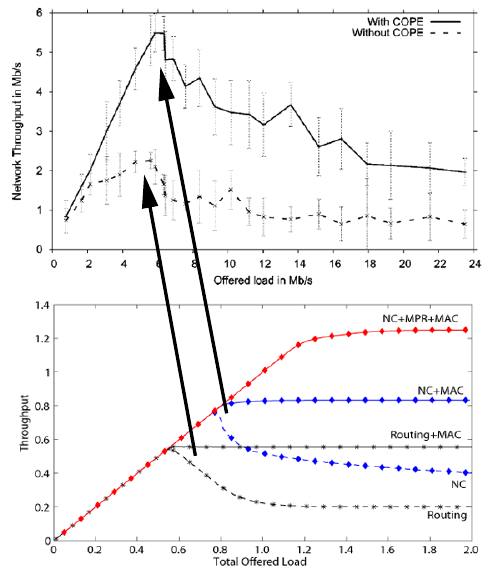}\vspace{-15pt}}
\par\end{centering}

\textcolor{black}{\caption{Comparison of the empirical COPE performance data collected from a
20-node 802.11 wireless ad hoc network test bed (top), \citep{Katti00},
and the resulting throughput using a model of the 802.11 MAC and 5-node
cross topology component proposed by \citep{Zhao00} (bottom). We
start with these elements to develop a model that can be used for
cross-layer design of various network technologies. The NC+MPR+MAC,
NC+MAC, and Routing+MAC curves are derived from our model and provide
an estimate of the achievable performance gains.\label{fig:Emperical-COPE-performance}}
\vspace{-20pt}}
\end{figure}

The development of the opportunistic inter-session NC scheme, COPE,
by Katti \textit{et. al.} \citep{Katti00} led to various models and
analyses that attempted to explain COPE's experimental results. Le
\textit{et. al.} \citep{4509678} and Sengupta \textit{et. al.} \citep{4215706}
developed models to describe these results, but only considered coding
a maximum of two packets together at a time or did not address the
interaction between NC and MAC fairness. \textcolor{black}{As a result,
their models provide throughput gains that are considerably smaller
than the experimental results and do not explain the non-monotonic
behavior shown in the upper half of Fig. \ref{fig:Emperical-COPE-performance}.
Zhao and Médard \citep{Zhao00} showed that the }\textit{\textcolor{black}{fairness}}\textcolor{black}{{}
imposed by the 802.11 MAC helps to explain this non-monotonic behavior.
In addition, they demonstrated that the majority of the throughput
gain achieved from COPE is a result of coding three or more unencoded,
or native, packets together at time. Their analysis showed that these
gains are not reflected in three node network models and that at least
five nodes are required to accurately capture the effects of COPE.
Fig. \ref{fig:Emperical-COPE-performance} shows that the 802.11 model
and 5-node cross topology component from \citep{Zhao00} is consistent
with the results from \citep{Katti00}. Furthermore, Seferoglu }\textit{\textcolor{black}{et.
al.}}\textcolor{black}{{} \citep{5487689} has used similar 5-node topology
components, and variants of them, to analyze TCP performance over
coded wireless networks. With this in mind, we develop our model using
the basic 5-node cross components from \citep{Zhao00} and \citep{5487689}
and other possible combinations of these 5-node topology components
in order to help in our understanding of the effects of combining
}NC\textcolor{black}{{} and various MAC implementations in larger networks.}

Finally, we demonstrate the capability of the model to predict the
gains and highlight design challenges from incorporating additional
technologies, such as MPR, into the network. MPR is known to enhance
wireless network performance and has been extensively researched with
unencoded traffic \textcolor{black}{\citep{1180550,1272,5196676,Tong01};
but little on the joint use of MPR, }NC\textcolor{black}{, and MAC
design exists. Garcia-Luna-Aceves}\textit{\textcolor{black}{{} et. al.}}\textcolor{black}{{}
\citep{1280960} }\textit{\textcolor{black}{compared}}\textcolor{black}{{}
the use of }NC\textcolor{black}{{} to MPR, but did not consider their
}\textit{\textcolor{black}{combined}}\textcolor{black}{{} use; and Rezaee
}\textit{\textcolor{black}{et. al.}}\textcolor{black}{{} \citep{Rezaee(00)}
provided an analysis of the combined use of }NC\textcolor{black}{{}
and MPR in a fully connected network, but did not consider the effects
of bottlenecks or multi-hop traffic. We show that our model provides
an intuitive method for determining an estimate of the achievable
gains from the combined use of MPR and }NC\textcolor{black}{{} in a
congested, multi-hop network.}

\textcolor{black}{The remainder of the paper is organized as follows:
Section \ref{sec:Network-models} provides a detailed description
of the network model; Section \ref{sec:Multi-Packet-Reception-and}
provides an example of }NC\textcolor{black}{{} and MPR for 5-node network
topology components using the existing 802.11 MAC; Section \ref{sec:Improving-the-MAC}
demonstrates the flexibility of using the model when considering the
design of various network elements; and we conclude in Section \ref{sec:Conclusion}.}

\section{\textcolor{black}{Network models and parameters\label{sec:Network-models}}}

Our main goal is to develop a simple model that gives insight into
cross-layer design of wireless networks by using NC, various MAC approaches,
and MPR as examples. To this end, we identify the fundamental behavior
of each aspect of the network and model each element using simple,
intuitive methods so that we can evaluate the potential throughput
gains.

\textcolor{black}{The performance of }NC\textcolor{black}{{} is modeled
by considering the ability of a given node to combine multiple packets
together as well as the primary implementation details of the particular
}NC\textcolor{black}{{} scheme. We use} COPE \citep{Katti00} as a case
study. COPE uses the broadcast nature of the wireless channel to opportunistically
code packets from different nodes together using a simple XOR operation.\textcolor{black}{{}
Any node that receives an encoded message is able to decode it using
the unencoded, or native, packets captured from the wireless channel.
We model COPE such that packet transmissions are never delayed. If
a node does not have more than one packet to encode, it does not wait
for another packet to arrive. Rather, it sends the packet unencoded
at the first opportunity. In addition, all packets headed towards
the same next-hop will not be encoded together because the next-hop
would not be able to decode these coded packets due to the lack of
enough degrees of freedom. We do not consider the complexity of the
coding or decoding operations nor any other aspects of the }NC\textcolor{black}{{}
implementation since their contributions to the overall network performance
is small in relation to the specific implementation aspects mentioned
above.}

\textcolor{black}{The MAC is modeled by identifying its primary behavior
in the network and simplifications are made by assuming optimal performance
from its secondary and tertiary behaviors. This gives us an intuitive
approach in determining the potential throughput while ensuring that
we understand the fundamental characteristics of the network. For
instance, the 802.11 MAC employs a fairness mechanism that distributes
channel resources equally among all competing nodes within a network.
As we show in the following sections, this is the primary cause of
the non-monotonic saturation behavior in the experimental throughput
shown in the upper half of Fig. \ref{fig:Emperical-COPE-performance}.
Since the fairness mechanism is the major contributor to overall network
performance, we assume optimal performance from all other aspects
of the MAC. For example, the non-monotonic behavior is a result of
both collisions and fairness; but the total effects of collisions
from either hidden nodes or identical back-off times on throughput
are small in relation to the effects of the 802.11 MAC fairness mechanisms.
Furthermore, we do not consider the additional effects on overall
throughput associated with the virtual 802.11 CS mechanisms (RTS/CTS)
or other aspects of the MAC such as the potential of lost channel
resources due to the MAC's random back-off. Instead, we assume optimal
performance from each of these secondary and tertiary behaviors. These
assumptions, as a result, provide upper bounds to the achievable throughput
in the various networks that employ the MAC.}

\textcolor{black}{We also show how additional techniques to increase
network performance, such as MPR, can be similarly modeled. MPR can
be implemented in a variety of methods from Code Division Multiple
Access (CDMA) or multiple-input-multiple-output (MIMO) to orthogonal
frequency division multiple access (OFDMA). In subsequent sections,
we model MPR by allowing each node to successfully receive $m$ multiple
packets at the same time.}

\begin{figure}
\begin{centering}
\textcolor{black}{\includegraphics[width=0.75\columnwidth]{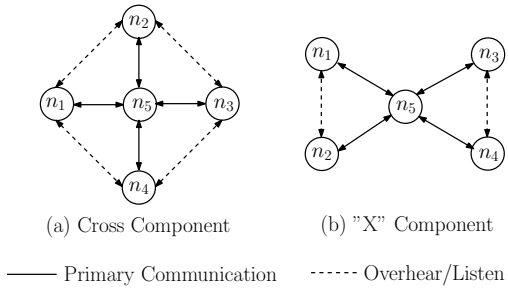}\vspace{-15pt}}
\par\end{centering}

\textcolor{black}{\caption{\label{fig:Network-Topologies}\textcolor{black}{Two of the basic
network structures responsible for traffic bottlenecks and congestion
in larger networks. All nodes are sources and all flows originating
from $n_{j}$, $j\in[1,4]$ cross at $n_{5}$.}}
\vspace{-20pt}}
\end{figure}

\textcolor{black}{Finally, our model uses several basic canonical
topology components that contain only five nodes where each node is
both a source and a sink. Two of the possible components are shown
in Fig. \ref{fig:Network-Topologies}. These components are of interest
because they form the primary structures in larger networks that create
bottlenecks and congestion. By looking at the traffic that travels
through the center node, these components help us model the performance
gains of multi-hop traffic under both low and high loads. While all
possible combinations of these basic canonical topology components
should be evaluated, we only focus on two of the components, shown
in Fig. \ref{fig:Network-Topologies}, since the analysis of the others
are redundant and provide little to the clarification of our approach.}

\textcolor{black}{Each component has specific constraints due to their
structure and will effect the performance of the MAC, }NC\textcolor{black}{,
and MPR in different ways. The center node $n_{5}$ in each component
is fully connected regardless of the topology, and traffic flows originating
from the center require only a single hop to reach their destination.
Within the cross topology component, each traffic flow originating
from a given node is terminated at the node directly opposite the
center; and in the ``X'' topology component, all flows originating
from a node in a given set terminates at a node in the opposite set.
Therefore, each flow must pass through the center regardless of topology.
For example, nodes $n_{1}$, $n_{2}$, and $n_{5}$ in the {}``X''
topology component are fully connected and nodes $n_{3}$, $n_{4}$,
and $n_{5}$ are also fully connected; but $n_{1}$ and $n_{2}$ are
not connected to $n_{3}$ and $n_{4}$. All traffic between any node
$\{n_{1},n_{2}\}\in X_{1}$ and any node $\{n_{3},n_{4}\}\in X_{2}$
must travel through the center.}

\textcolor{black}{In an effort to simplify the following explanation,
we make several additional simplifications which can be easily incorporated
into the model. We assume feedback is perfect, the load required for
acknowledgments are contained as part of the initial transmission's
load, and the wireless channel is loss-less. We also consider the
additional constraint that each node is half-duplex, i.e. a node will
overhear any transmission from its neighbors only if it is not transmitting.
When considering the model for MPR, we allow $m$ packets to be sent
from different sources in a single time slot with the constraint that
we try to maximize the number of neighbors a node can overhear within
any given time slot. In essence, we duplicate CSMA/CA for each $m=2$
case in the sense that a node will transmit only if none of its neighbors
are transmitting. For cases where $m\geq3$, we pick the combination
of transmitting nodes such that the average number of transmissions
received by any given node within the network is maximized.}

\textcolor{black}{Sections \ref{sec:Multi-Packet-Reception-and} and
\ref{sec:Improving-the-MAC} provide both an analysis of the maximum
achievable throughput and simulations over various values of the total
offered load $P$ to the network. The load $P$ to the network from
the set of source nodes $i\in\mathcal{N}$, is defined as $P=\sum_{i\in\mathcal{N}}\rho_{i}$,
where $\mathcal{N}$ is the set of nodes in the topology component
and $\rho_{i}=\nicefrac{k_{i}}{100}$ is node $i$'s individual load
contribution, or the fraction of time required to send all of its
$k_{i}$ packets to the next-hop. We stochastically determine $k_{i}$
using a binomial distribution given $P$ with parameters $n_{i}=K-\sum_{j=1}^{i-1}k_{j}$
and $p_{i}=\nicefrac{1}{(N-i+1)}$, $N=\left|\mathcal{N}\right|$,
in each iteration of our simulation and average these results for
each total offered load evaluated. We then use the model described
in this section to determine the total network throughput $S$ which
is equivalent to the total number of packets divided by number of
time slots needed to send every packet to its intended destination.}

\section{\textcolor{black}{Multi-Packet Reception and Network Coding Performance
Analysis\label{sec:Multi-Packet-Reception-and}}}

\textcolor{black}{With each of the network topology components shown
in Fig. \ref{fig:Network-Topologies}, we provide an example of the
performance analysis with and without the use of }NC\textcolor{black}{{}
and MPR when the 802.11 MAC is used. We also consider both unicast
and broadcast traffic where a unicast transmission is complete when
all packets from each source successfully reaches their destination
and a broadcast transmission is complete when all nodes have received
every packet from all sources.}

\subsection{\textcolor{black}{Cross Topology Component Analysis\label{sub:Cross-Network-Topology}}}

\textcolor{black}{Each node $i\in[1,5]$, requires $\rho_{i}$ of
the time to send all of its packets one hop. The resulting total offered
load to the network is then $P=\sum_{i=1}^{5}\rho_{i}$. We define
the total network component load $P_{T}=\rho_{R}+\rho_{M}$ as the
time required to send all packets through the topology component where
$\rho_{R}$ is the required load to relay packets and $\rho_{M}$
is the required load to send each native packet one hop. The load
required to relay packets through the center is $\rho_{R}=\frac{1}{c}\sum_{j=1}^{4}\rho_{j}$
where $c$ is the number of packets that can be encoded together which
is dependent on the number of neighbors a given node can overhear.
In the case of the cross topology component and enough packets to
code together, $c=4$ for $m=\{1,2\}$ and $c=2$ for all $m\geq3$.
The load needed to send each node's unencoded packets one hop is defined
as $\rho_{M}\geq\nicefrac{1}{m}\sum_{j=1}^{4}\rho_{j}+\rho_{5}$.
Because we setup the model so that we maximize the number of neighbors
any given node can overhear when using MPR, this expression is met
with equality if each $\rho_{j}$, $j\in[1,4]$, is equal. Otherwise,
the load $\rho_{M}$ is lower bounded by this equation and is a function
that is dependent on both the topology component's configuration and
each node's load imbalance. Finally, let the fraction of allocated
time slots a node receives as a result of the MAC be $s_{i}$.}

\begin{figure}
\begin{centering}
\textcolor{black}{\includegraphics[width=3.25in]{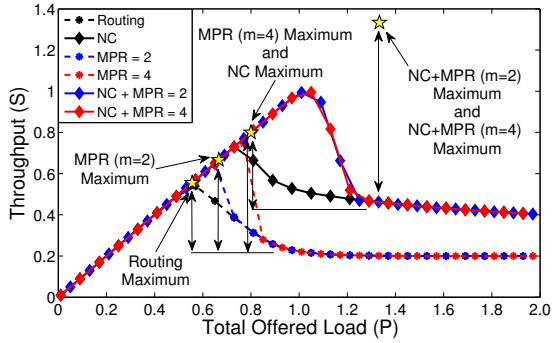}\vspace{-15pt}}
\par\end{centering}

\textcolor{black}{\caption{\label{fig:Cross-OriginalMAC}Average unicast and broadcast throughput
for a 5-node cross topology component. Each vertical double arrow
shows the difference in the maximum and saturated throughput due to
MAC fairness for each case.}
\vspace{-15pt}}
\end{figure}

\textcolor{black}{The throughput $S$ with unicast and broadcast traffic
is shown as a function of $P$ in Fig. \ref{fig:Cross-OriginalMAC}.
Each curve is obtained through simulation and is an average over the
load distribution discussed in Section \ref{sec:Network-models}.
Each star is obtained by analysis and}\textcolor{blue}{{} }\textcolor{black}{depicts
the maximum achievable throughput when the MPR and/or }NC\textcolor{black}{{}
gain is maximized. When $P_{T}<1$, each node is allocated enough
time slots to send all of its packets, and the allocated load is $s_{j}=\rho_{j}$
for $j\in[1,4]$ and $s_{5}=\rho_{5}+\rho_{R}$. The throughput $S$
increases linearly as the network load increases, regardless of the
use of MPR or }NC\textcolor{black}{, and reaches a maximum for each
case when $P_{T}=1$. The throughput then transitions into a saturated
region for $P_{T}>1$, where the allocated load for each node is $s_{j}\leq\rho_{j}$
and $s_{5}\leq\rho_{5}+\rho_{R}$. When }NC\textcolor{black}{{} is not
used, the throughput is $S=s_{5}$; and when }NC\textcolor{black}{{}
is used, the throughput will be a function of the number of packets
that can be encoded together.}

\subsubsection{\textcolor{black}{Routing (No Network Coding, $m=1$)}}

\textcolor{black}{We use routing as the performance baseline. Consistent
with the results found in \citep{Katti00} and the analysis performed
in \citep{Zhao00}, the throughput increases linearly within the non-saturated
region, $P\in[0,\nicefrac{5}{9})$. At $P=\nicefrac{5}{9}$, the throughput
reaches a maximum of $\nicefrac{5}{9}$ (depicted by a star in Fig.
\ref{fig:Cross-OriginalMAC}). This occurs when each sources' load
reaches $\rho_{i}=\nicefrac{1}{9}$ for $i\in[1,5]$. The total load
of the center node, as a consequence, is $\rho_{5}+\rho_{R}$ where
$\rho_{R}=\sum_{i=1}^{4}\rho_{j}=\nicefrac{4}{9}$. Since $P_{T}=1$,
$s_{j}=\rho_{j}$ and $s_{5}=\rho_{5}+\rho_{R}$.}

\textcolor{black}{The throughput saturates for $P>\nicefrac{5}{9}$.
Initially, the 802.11 MAC allocates time slots to nodes requiring
more resources. The throughput is therefore the amount of time $n_{5}$
is able to transmit, $s_{5}=1-\sum_{i=1}^{4}s_{i}$, which decreases
as $P$ increases. The network component completely saturates when
each node requires a large fraction of the available time slots but
the MAC restricts each node's access to the channel by ensuring fairness
among each competing node, i.e. $s_{i}=\nicefrac{1}{5}$ for $i\in[1,5]$.
For large enough $P$, the throughput saturates to the total amount
of information that $n_{5}$ can transmit, i.e., $S=s_{5}=\nicefrac{1}{5}$.}

\subsubsection{\textcolor{black}{Network Coding Only ($m=1$)}}

\textcolor{black}{We now allow }NC\textcolor{black}{{} to be used by
the center node. Each node transmits one at a time, allowing each
node to receive four native packets or four degrees of freedom (three
degrees of freedom through the use of opportunistic listening plus
one degree of freedom from the packet originating at the node). After
each edge node has completed transmission, node $n_{5}$ transmits
a single encoded packet, which is sufficient for each edge node to
obtain the single degree of freedom it still requires to complete
both the unicast and broadcast sessions.}

\textcolor{black}{From Fig. \ref{fig:Cross-OriginalMAC}, when $P\in[0,\nicefrac{5}{9})$,
}NC\textcolor{black}{{} is seen to provide no additional gains over
the use of routing alone since $n_{5}$ can forward each packet received
without the MAC limiting its channel use. For $P\in[\nicefrac{5}{9},\nicefrac{5}{6}),$
}NC\textcolor{black}{{} is instrumental in achieving the throughput
shown. The MAC does not limit channel resources until the maximum
throughput of $S=\nicefrac{5}{6}$ is reached when $P_{T}=\sum_{i=1}^{5}\rho_{i}+\frac{1}{4}\sum_{j=1}^{4}\rho_{j}=1$
where $\rho_{i}=\nicefrac{1}{6}$ for $i\in[1,5]$. At this maximum,
the MAC ensures fairness among all competing nodes and the throughput
saturates. As $P$ increases, the gain provided by }NC\textcolor{black}{{}
diminishes. The number of packets reaching $n_{5}$ from each edge
node is limited by the MAC while packets introduced into the network
component by $n_{5}$ are not. The coding gain, therefore, approaches
zero as $P\rightarrow\infty$.}

\subsubsection{\textcolor{black}{Multi-Packet Reception of Order 2 and 4 (No Network
Coding and $m=\{2,4\}$)}}

\textcolor{black}{MPR is similar to the routing case described earlier
except we now allow a maximum of $m$ edge nodes to transmit within
a given time slot. For $m=2$, the total time used by all of the edge
nodes to transmit their packets to $n_{5}$ is $\nicefrac{1}{2}$
that needed by routing while the center node cannot transmit multiple
packets simultaneously and must transmit each received packet individually.
Using CSMA, which restricts nodes opposite each other to transmit
at the same time, the point at which the protocol saturates for symmetric
source loads occurs when $P_{T}=\sum_{j=1}^{4}\rho_{j}+\nicefrac{1}{2}\sum_{j=1}^{4}\rho_{j}+\rho_{5}=1$
where $\rho_{i}=\nicefrac{1}{7}$ for $i\in[1,5]$. This yields the
maximum throughput of $S=\nicefrac{5}{7}$. The throughput saturates
to the same throughput as routing for values of $P_{T}>1$ and the
gain for $m=2$ in the saturated regime is 1 due to the suboptimal
saturation behavior of the protocol.}

\textcolor{black}{The behavior for $m=4$ is the same as that for
$m=2$ except the maximum of $S=\nicefrac{5}{6}$ occurs when $P_{T}=\sum_{j=1}^{4}\rho_{j}+\nicefrac{1}{4}\sum_{j=1}^{4}\rho_{j}+\rho_{5}=1$
where $\rho_{i}=\nicefrac{1}{6}$. We allow all edge nodes to transmit
their packets to $n_{5}$ simultaneously, requiring a total of $\nicefrac{1}{6}$
of the time-slots. Node $n_{5}$ then sends each node's packet individually,
including its own, to the intended recipient requiring the remainder
of the time slots to finish each unicast/broadcast transmission. As
$P$ increases, the MAC limits each node's number of available time
slots and $S$ saturates to $\nicefrac{1}{5}$. Again, the gain in
the saturated region for $m=4$ is equal to the cases of $m=2$ and
routing.}

\subsubsection{\textcolor{black}{Network Coding with Multi-Packet Reception of Order
2 and 4 ($m=2,4$)}}

\textcolor{black}{For $m=2$, the maximum throughput of $S=\nicefrac{5}{4}$
occurs when $P_{T}=\nicefrac{1}{4}\sum_{j=1}^{4}\rho_{j}+\nicefrac{1}{2}\sum_{j=1}^{4}\rho_{j}+\rho_{5}=1$
where $\rho_{i}=\nicefrac{1}{4}$ for $i\in[1,5]$. Each set of nodes,
$\{n_{1},n_{3}\}$ and $\{n_{2},n_{4}\}$, uses $\nicefrac{1}{4}$
of the total number of time slots to transmit to $n_{5}$ which then
transmits a single encoded packet derived from all four node's native
packets in addition to its own native packet. For $P_{T}>1$, the
throughput saturates to the saturated }NC\textcolor{black}{{} throughput
due to the 802.11 MAC. While the maximum achievable throughput is
$\nicefrac{25}{16}$ times the }NC\textcolor{black}{{} without MPR throughput,
the saturated gain for $m=2$ is equal to the gain found when }NC\textcolor{black}{{}
was used alone in this region.}

\textcolor{black}{The throughput using }NC\textcolor{black}{{} and $m=4$
for unicast traffic is equivalent to }NC\textcolor{black}{{} and $m=2$.
All four edge nodes transmit to $n_{5}$ which then transmits two
encoded packets in addition to its own; or we limit the number of
simultaneous transmissions to two thus allowing $n_{5}$ to code everything
together and send a single encoded packet to all of the edge nodes.
Either strategy will achieve the same gain although the difference
occurs when considering either unicast (former option) or broadcast
(later option). The maximum throughput for broadcast traffic using
the first method is $S=1$, and $S=\nicefrac{5}{4}$ for the second
which is consistent with the maximum unicast throughput. This difference
indicates that increasing $m$ when using }NC\textcolor{black}{{} may
not be the optimal strategy. Although we do not show it here, the
canonical topology components can be easily modified to include any
number of nodes which would allow us to further look into the optimal
strategy for broadcast traffic.}

\subsection{\textcolor{black}{{}``X'' Topology Component\label{sub:X-Topology-Component}}}

\textcolor{black}{}
\begin{figure}
\centering{}\textcolor{black}{\includegraphics[width=3.25in]{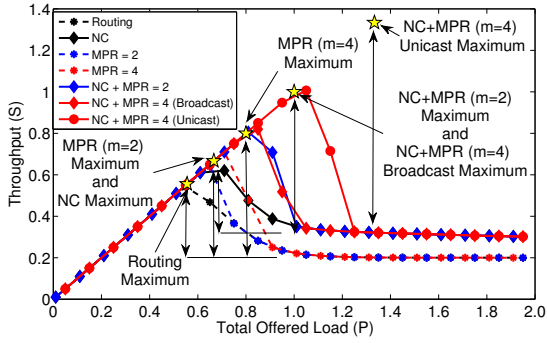}\vspace{-5pt}\caption{\label{fig:X-OriginalMAC}Average broadcast and unicast throughput
for a 5-node {}``X'' topology component. Each vertical double arrow
shows the difference in the maximum and saturated throughput due to
MAC fairness for each case.}
\vspace{-10pt}}
\end{figure}
\textcolor{black}{The cross topology component gives insight into
the performance of COPE and MPR in a dense network, and it represents
the best case scenario when COPE is used since it maximizes the number
of transmissions any given node receives. In order to understand the
behavior of COPE and MPR in sparser networks, we limit the number
of each node's neighbors by analyzing the behavior of COPE and MPR
in the {}``X'' topology component shown Fig. \ref{fig:Network-Topologies}(b).
Fig. \ref{fig:X-OriginalMAC} shows both the maximum and average throughput
resulting from the use of the {}``X'' topology component. Within
this section, we focus only on the cases involving }NC\textcolor{black}{{}
since it can be easily verified that the routing and $m=\{2,4\}$
analysis for this topology component are the same as the cross topology
component analysis.}

\subsubsection{\textcolor{black}{Network Coding Only ($m=1$)}}

\textcolor{black}{Limiting the ability to overhear other edge nodes
in the component results in the reduction in the number of possible
packets that can be encoded together. Packets from different nodes
within the same set, i.e. $\{n_{1},n_{2}\}\in X_{1}$ and $\{n_{3},n_{4}\}\in X_{2}$,
cannot be encoded together because all flows transitioning between
$X_{1}$ and $X_{2}$ are effectively headed towards the same next-hop.
This forces $n_{5}$ to code only a subset of packets together which
increases the number of transmissions the center node must make. For
example, the center node must make a minimum of two transmissions
for every four packets it receives from different edge nodes in order
to ensure that each node has the necessary degrees of freedom to decode
all of the packets.}

\textcolor{black}{Like the cross component's throughput, the throughput
of the {}``X'' topology component increases linearly until it reaches
its maximum at $S=\nicefrac{5}{7}$. Assuming symmetric source loads,
this maximum occurs when $P_{T}=\nicefrac{1}{2}\sum_{j=1}^{4}\rho_{j}+\sum_{j=1}^{5}\rho_{j}=1$
where $\rho_{i}=\nicefrac{1}{7}$ for $i\in[1,5]$. The throughput
saturates for $P_{T}>1$ and the non-monotonic behavior in the saturated
throughput regime is again due to the fairness aspect of the 802.11
MAC. By comparing the results obtained from both the cross and {}``X''
topology components, it is evident that the performance of COPE is
highly dependent on the network structure. As the network becomes
sparser, the gain from COPE is diminished.}

\subsubsection{\textcolor{black}{Network Coding with Multi-Packet Reception of Order
2 and 4 ($m=\{2,4\}$)}}

\textcolor{black}{For $m=2$, the throughput increases linearly until
it reaches its maximum at $S=1$ when $P_{T}=\nicefrac{1}{2}\sum_{j=1}^{4}\rho_{j}+\nicefrac{1}{2}\sum_{j=1}^{4}\rho_{j}+\rho_{5}=1$
where $\rho_{i}=\nicefrac{1}{5}$ for $i\in[1,5]$. The throughput
then saturates to the }NC\textcolor{black}{{} throughput for $P_{T}>1$.
For $m=4$, the maximum unicast throughput is $S=\nicefrac{5}{4}$
and is achieved when $P_{T}=\nicefrac{1}{2}\sum_{j=1}^{4}\rho_{j}+\nicefrac{1}{4}\sum_{j=1}^{4}\rho_{j}+\rho_{5}=1$
where $\rho_{i}=\nicefrac{1}{4}$ for $i\in[1,5]$. The center codes
a maximum of two packets together from different edge node sets and
transmits two encoded packets back to the set of edge nodes in addition
to its own native packet. This gives each edge node enough degrees
of freedom to complete all unicast transmissions. When considering
broadcast traffic, each node still requires a maximum of one additional
degree of freedom. Allowing $n_{5}$ to code all of the edge node's
native packets together and send one encoded transmission enables
each node to extract the required degree of freedom and obtain the
full set of messages from each source. The maximum throughput is therefore
the same as the case for }NC\textcolor{black}{{} with $m=2$ and is
equal to $S=1$. Similar to the cross topology component, the average
throughput for both cases discussed in this section does not reach
the maxima found because of the stochastic load distribution, which
results in asymmetric traffic flows across the center node that limits
the effectiveness of both COPE and the implementation of MPR that
we chose. If each node has an equal amount of information to send,
the maxima found in this section would be achieved.}

\begin{figure}
\begin{centering}
\textcolor{black}{\includegraphics[width=3.2in]{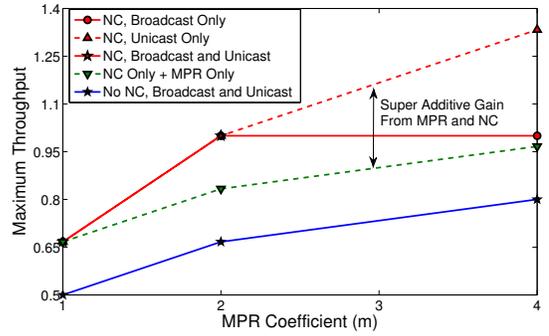}\vspace{-15pt}}
\par\end{centering}

\textcolor{black}{\caption{\textcolor{black}{Maximum throughput of the 5-node {}``X'' topology
component as function of the MPR capability. Super-additive gains
are achieved when using }NC\textcolor{black}{{} in conjunction with
MPR. This is shown by comparing the throughput obtained using both
}NC\textcolor{black}{{} and MPR with the throughput that would be obtained
if the gain from using }NC\textcolor{black}{{} alone is added with the
gain obtained from using MPR alone (NC Only + MPR Only).\label{fig:sat-flow}}}
\vspace{-20pt}}
\end{figure}

The use of these components within our model allows us to determine
the fundamental behavior of combining COPE and MPR in a larger network.\textcolor{black}{{}
For example, Fig. \ref{fig:sat-flow} shows a summary of our analysis
by plotting the maximum unicast and broadcast throughput as a function
of the MPR capability. It shows the super-additive throughput behavior
when MPR is used in conjunction with }NC\textcolor{black}{{} by comparing
this throughput with the throughput that would be obtained by adding
the individual gains obtained using MPR and }NC\textcolor{black}{{}
separately.}

\textcolor{black}{The use of the {}``X'' topology component allows
us to determine behavior that was otherwise masked in the cross component.
It provided insight into the behavior of }NC\textcolor{black}{{} and
MPR in sparser networks, methods in which to implement variants of
COPE for broadcast traffic, and highlighted the super-additive gains
from combining the two communication technologies. We can also look
into variants of the cross and {}``X'' components; but they are
not discussed since they provide little clarification to the presentation
of the model. In this example, however, these variants provide insight
into the robustness of MPR and COPE's throughput gains to topology
changes.}

\section{\textcolor{black}{Improving the MAC Fairness Protocol\label{sec:Improving-the-MAC}}}

The use of our model highlighted a major drawback to the use of the
802.11 MAC in multi-hop networks. The non-monotonic throughput behavior
that is evident in Fig. \ref{fig:Cross-OriginalMAC} and \ref{fig:X-OriginalMAC}
significantly reduces throughput as the offered load to the network
increases. We now show how the model can be used to redesign the MAC
so that we eliminate this sub-optimal behavior in the presence of
NC, MPR, and their combination. Furthermore, we use the model to develop
a MAC approach that provides fairness to \textit{flows} rather than
to \textit{nodes}.

\textcolor{black}{It is obvious from Section \ref{sec:Multi-Packet-Reception-and}
that if we want to eliminate the non-monotonic saturation throughput
behavior, we want to allocate more channel resources to the center
node than the edge nodes. We choose to allocate resources proportional
to the amount of non-self-generated traffic flowing through each node
when the network saturates. While allocating fewer resources to flows
originating at the center and more resources to flows originated at
edge nodes yields even higher throughput, this approach ensures that
each flow of information is given the same priority. The center node
will be allocated more resources than each edge node in order to relay
information; but it must also limit the amount of self-generated traffic
so that it equals the average per-node non-self-generated traffic
being relayed.}

\textcolor{black}{We design the revised MAC using a slight modification
of the components found in Fig. \ref{fig:Network-Topologies}. For
the cross topology component, we let there be $N-1$ edge nodes and
a single center, or relay, node. All edge nodes are connected with
the center node and connected with all other edge nodes except the
one directly opposite the center. For the {}``X'' topology component,
we also let there be $N-1$ edge nodes and a single center node. The
edge nodes are split into two sets $X_{1}$ and $X_{2}$. All edge
nodes within a given set are fully connected and also connected to
the center node. Within the cross topology component, each node communicates
with the node directly opposite the center. In the {}``X'' topology
component, each node communicates with a node in a different set.}

\textcolor{black}{The allocated number of time slots each node receives
so that the throughput $S$ is maximized, subject to the flow constraints
and $\sum_{j=1}^{N-1}\nicefrac{s_{j}}{m}+s_{R}=1$, is divided into
three cases where $s_{j}$ is the fraction of time slots allocated
to each edge node and $s_{R}$ is the fraction of time slots allocated
to the center node. Similar to Section \ref{sec:Multi-Packet-Reception-and},
the throughput $S=s_{R}$ when }NC\textcolor{black}{{} is not used.
When }NC\textcolor{black}{{} is used, $S$ is a function of the number
of packets that can be coded together, which is dependent on the density
of the network, the MPR coefficient $m$, the use of CSMA, and the
traffic type (unicast or broadcast). In order to simplify the explanation
within this paper, we limit the examples we explore by considering
only values of $m=\{1,2,4\}$ and symmetric source loads. In the case
of the {}``X'' topology component, we also restrict our example
to situations where the cardinality of each set $X_{1}$ and $X_{2}$
are equal. The cases include:}

\textsl{\textcolor{black}{Cross Topology Component with Unicast Traffic
or Broadcast Traffic:}}\textcolor{black}{{} The cross topology component
can be used to design part of the MAC for operation in dense networks.
Without }NC\textcolor{black}{, the center node requires a number of
time slots equal to the number of source nodes $N$. With }NC\textcolor{black}{,
throughput is maximized by ensuring the center node codes the maximum
number of native packets together. Implementing MPR for $m=4$ can
potentially prevent each node from immediately decoding any coded
message sent by the center. This is due to the potential of a given
node transmitting at the same time as one of its neighbors. Generalizing
for $N$ and $m$ as well as considering only integer numbers of time
slots:
\begin{equation}
s_{j}=\begin{cases}
\frac{1}{\lceil(N-1)/m\rceil+N} & \textrm{without NC}\\
\frac{1}{\lceil(N-1)/m\rceil+m} & \textrm{with NC}
\end{cases}\label{eq:cross_bw1}
\end{equation}
and
\begin{equation}
s_{R}=\begin{cases}
\frac{N}{\lceil(N-1)/m\rceil+N} & \textrm{without NC}\\
\frac{m}{\lceil(N-1)/m\rceil+m} & \textrm{with NC}
\end{cases}\label{eq:cross_bw2}
\end{equation}
}

\textsl{\textcolor{black}{{}``X'' Topology Component: }}\textcolor{black}{Using
the {}``X'' topology component helps gain insight into the design
of the MAC for operation in sparser networks. From the 802.11 example
in Section \ref{sub:X-Topology-Component}, we determined that the
throughput differs for both unicast and broadcast traffic. As a result,
we define the fraction of time slots $s^{U}$ allocated to each node
for unicast traffic as:
\begin{equation}
s_{j}=s_{j}^{U}=\begin{cases}
\frac{1}{\lceil(N-1)/m\rceil+N} & \textrm{without NC}\\
\frac{1}{\lceil(N-1)/m\rceil+\max\left(\mid X_{1}\mid,\mid X_{2}\mid\right)+1} & \textrm{with NC}
\end{cases}\label{eq:x_bw1}
\end{equation}
and 
\begin{equation}
s_{R}=s_{R}^{U}=\begin{cases}
\frac{N}{\lceil(N-1)/m\rceil+N} & \textrm{without NC}\\
\frac{\max\left(\mid X_{1}\mid,\mid X_{2}\mid\right)+1}{\lceil(N-1)/m\rceil+\max\left(\mid X_{1}\mid,\mid X_{2}\mid\right)+1} & \textrm{with NC}
\end{cases}\label{eq:x_bw2}
\end{equation}
When considering broadcast traffic, additional degrees of freedom
may need to be sent by the center to complete the session. For $m=\{1,2\}$,
no additional degrees of freedom are required by any node. For the
case involving }NC\textcolor{black}{{} and $m=4$, each edge node will
require one additional degree of freedom in order to decode all of
the encoded packets sent by the center. As a result, the denominator
in the }NC\textcolor{black}{{} equations of (\ref{eq:x_bw1}) and (\ref{eq:x_bw2})
is replaced by $\left\lceil \nicefrac{(N-1)}{4}\right\rceil +\max(X_{1},X_{2})+2$,
as well as the numerator in the }NC\textcolor{black}{{} case of (\ref{eq:x_bw2})
with $\max(X_{1},X_{2})+2$.}

\begin{figure}
\centering{}\textcolor{black}{\includegraphics[width=3.25in]{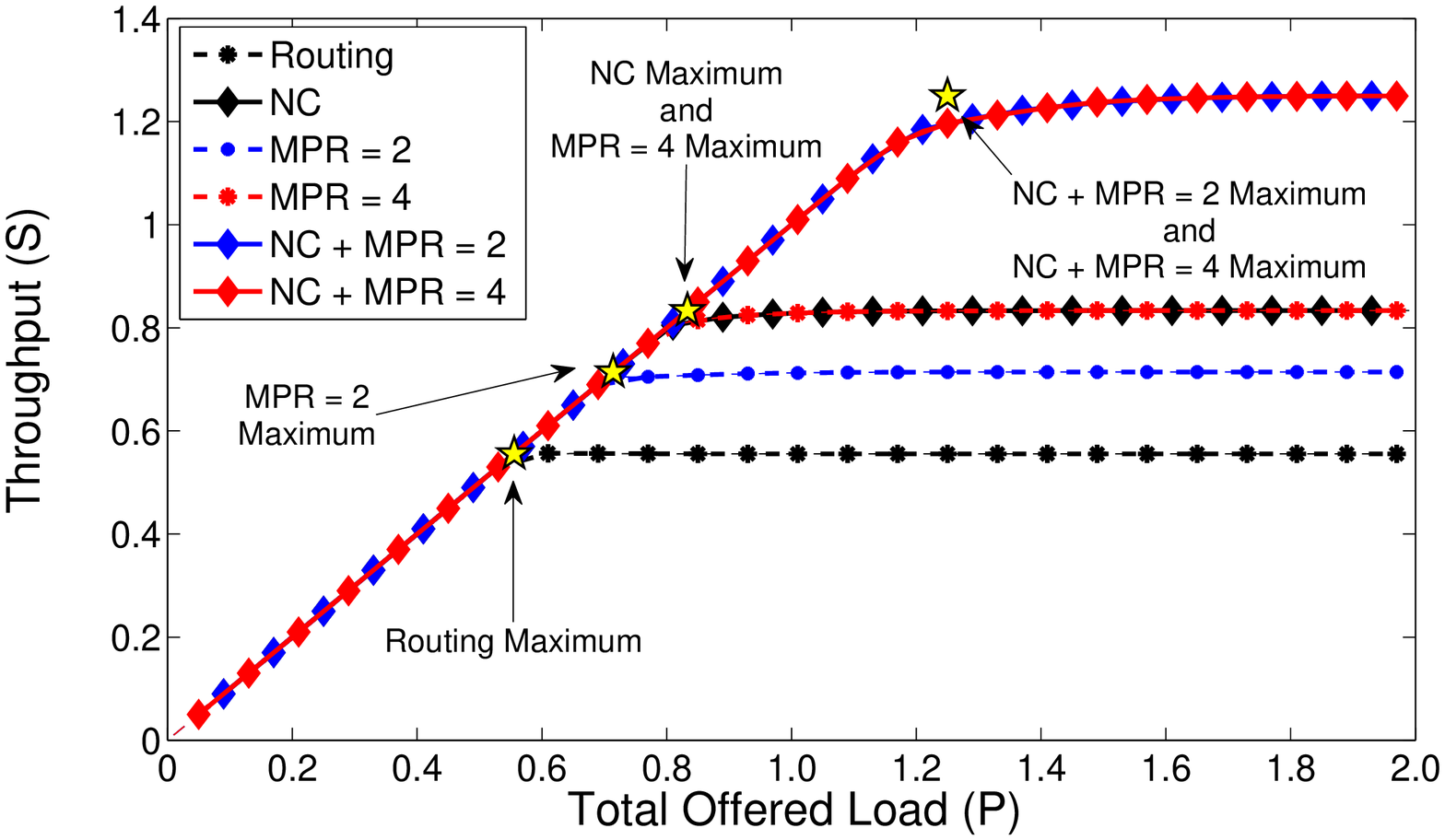}\vspace{-5pt}\caption{\label{fig:5Node-Cross_ImprovedMAC}5-Node cross topology component
unicast and broadcast throughput using the improved MAC.}
\vspace{-10pt}}
\end{figure}

\begin{figure}
\begin{centering}
\textcolor{black}{\includegraphics[width=3.2in]{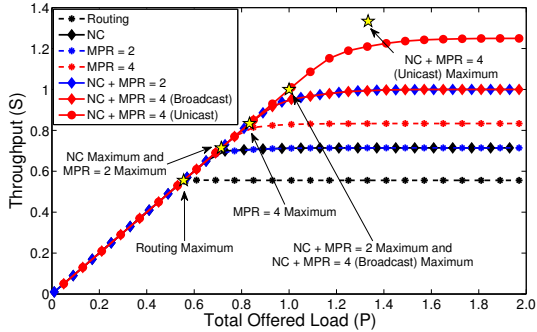}\vspace{-15pt}}
\par\end{centering}

\textcolor{black}{\caption{\label{fig:5-Node-X-Unicast}5-Node {}``X'' component throughput
using the improved MAC.}
\vspace{-10pt}}
\end{figure}

\textcolor{black}{We apply the revised fairness protocol to both the
5-node cross and {}``X'' topology components using the same methods
described in Section \ref{sec:Multi-Packet-Reception-and}. We find
that the throughput saturates at the }\textit{\textcolor{black}{maxima}}\textcolor{black}{{}
found in Section \ref{sec:Multi-Packet-Reception-and} for each topology
component. Fig. \ref{fig:5Node-Cross_ImprovedMAC} and \ref{fig:5-Node-X-Unicast}
show both the unicast and broadcast throughput for the cross and {}``X''
topology components, respectively, using our improved MAC approach.
It is clear by comparing Figures \ref{fig:Cross-OriginalMAC} and
\ref{fig:X-OriginalMAC} with Figures \ref{fig:5Node-Cross_ImprovedMAC}
and \ref{fig:5-Node-X-Unicast} respectively that the new MAC eliminates
the non-monotonic saturation behavior. Furthermore, this comparison
shows that the combination of }NC\textcolor{black}{, MPR with $m=\{2,4\}$,
and the new MAC provides throughput gains on the order of six times
that of routing alone with the 802.11 MAC in the saturated throughput
regime.}
\textcolor{black}{}

\section{\textcolor{black}{Conclusion\label{sec:Conclusion}}}

We developed a simple, intuitive model by approximating key network
elements with simple models of their underlying primary behavior.
We then used the model to evaluate the performance of specific implementations
of the 802.11 MAC with COPE and MPR in multi-hop networks. Gaining
key insight into design strategies for combining the three technologies,
each scenario presented gave a rough order of magnitude for the performance
of implementing the MAC, NC, MPR, and their combination in larger
networks. The model further shows that combining COPE with MPR results
in super-additive throughput gains. We then demonstrated that the
non-monotonic saturation experienced in \citep{Katti00} is explained
by the sub-optimal behavior of the 802.11 MAC, and used our model
to develop a MAC approach tailored toward the combined use of COPE
and MPR that provides monotonic saturation behavior, as well as fairness
to \emph{flows} rather than \emph{nodes}.\vspace{-15pt}

\textcolor{black}{\bibliographystyle{IEEEtran}
\bibliography{COPEMPRBib}
}
\end{document}